\documentclass[12pt]{article}
\usepackage{latexsym}
\usepackage{amsmath,amssymb,amsfonts,amsthm}
\usepackage[english]{babel}
\newcommand{\be}{\begin{equation}}
\newcommand{\ee}{\end{equation}}
\newcommand{\bea}{\begin{eqnarray}}
\newcommand{\eea}{\end{eqnarray}}
\newcommand{\nn}{\nonumber \\}
\newcommand{\p}[1]{(\ref{#1})}
\newcommand{\lb}{\label}
\def\theequation{\arabic{section}.\arabic{equation}}
\begin{document}
\begin{titlepage}

\vfill

\begin{center}
\baselineskip=16pt {\Large\bf Bispinor Auxiliary Fields in Duality-Invariant}
\vspace{0.2cm}

{\Large\bf Electrodynamics Revisited
}
\vskip 0.6cm {\large {\sl }} \vskip 10.mm {\bf E.A.
Ivanov, $\;$ B.M. Zupnik }
\vspace{1cm}

{\it Bogoliubov Laboratory of Theoretical Physics, JINR, \\
141980 Dubna, Moscow Region, Russia\\
}
\vspace{0.3cm}

{\tt eivanov@theor.jinr.ru},  $\;$ {\tt zupnik@theor.jinr.ru}
\end{center}
\vspace{1.6cm}

\par
\begin{center}
{\bf ABSTRACT}
\end{center}
\begin{quote}
Motivated by a recent progress in studying the duality-symmetric models of
nonlinear electrodynamics, we revert to the auxiliary tensorial (bispinor)
field formulation of the $O(2)$ duality proposed by us
in {\tt hep-th/0110074, 0303192}.
In this approach, the entire information about the given duality-symmetric
system is encoded in the $O(2)$-invariant interaction Lagrangian which
is a function of the auxiliary fields $V_{\alpha\beta},
\bar{V}_{\dot\alpha\dot\beta}\,$. We extend this setting to duality-symmetric
systems with higher derivatives and show that the recently employed
``nonlinear twisted self-duality constraints'' amount to the equations of
motion for the auxiliary tensorial fields in our approach. Some other
related issues are briefly discussed and a few instructive examples are
explicitly worked out.
\vspace{4.5cm}


\vfill \vfill \vfill \vfill \vfill
\end{quote}
\end{titlepage}

\setcounter{footnote}{0}

\setcounter{page}{1}

\section{Introduction}
A wide class  of the nonlinear electrodynamics models including the
renowned Born-Infeld (BI) theory exhibits  the on-shell $U(N)$ duality
invariance (or a less restrictive discrete self-duality) \cite{GZ}-\cite{AFZ}.
Recently, there was a rebirth of interest in the duality-invariant
theories, including those with higher derivatives (see, e.g.,
\cite{BHN}-\cite{RT2}), in connection with the problems of the
extended supergravity counterterms. There is the hope that the
duality symmetry considerations at the quantum level could play the
decisive role in checking the ambitious hypothesis of ultraviolet
finiteness of ${\cal N}=8, 4D$ supergravity (and/or its
higher-dimensional cousins and ${\cal N}< 8, 4D$
descendants)\cite{BHN,RKa,BN}.

About decade ago, we proposed a new general formulation of the
duality-invariant theories exploiting the tensorial (bispinor)
auxiliary fields \cite{IZ,IZ1,IZ2}. In this formulation, the $U(N)$ or discrete duality symmetries
acquire a nice interpretation as the {\it linear} off-shell
symmetries of the nonlinear interaction constructed out of the
auxiliary fields. These symmetries are broken in the total action,
since its free bilinear part is not invariant. The famous
Noether-Gaillard-Zumino (NGZ) constraint \cite{GZ2,GR} is linearized in this
formalism and, as a consequence, the duality symmetry between the
Bianchi identities and equations of motion becomes manifest. The
basic algebraic equations of motion relate the auxiliary fields to
the Maxwell field strengths and, in general, can be solved by some
recursive procedure. Substituting the solutions for the auxiliary
fields back into the original action, we obtain the perturbative
expansion of the nonlinear Lagrangians in terms of the
electromagnetic field strengths. In some particular cases of the
$U(1)$ duality invariant models,  the basic equations are reduced to
the algebraic equations of the orders 2, 3 or 4, and it becomes
possible to write their solutions (and the eventual actions) in a
closed form. In particular, the quadratic equation gives rise to the
notorious Born-Infeld (BI) theory.

One of the aims of the present paper is to argue that the methods
recently proposed in \cite{BN,CKR,CKO} for the systematic
construction of various duality-invariant nonlinear Lagrangians are
in fact equivalent to our approach \cite{IZ,IZ1,IZ2}. In particular,
``the nonlinear twisted duality constraint'' of \cite{BN,CKR,CKO} is
none other than the equation of motion for our bispinor auxiliary
field. Realizing this motivated us to return to the original
formulation in order to see how the latest developments in the
duality stuff can be inscribed into its framework.

In Section 2 we review our auxiliary-field formalism for the $U(1)$ (or $O(2)$)
duality symmetric electrodynamics models in the simplest case
without derivatives of the field strengths. We demonstrate that the
model with the simplest quartic auxiliary  $O(2)$ invariant
interaction $\frac12V^2\bar{V}^2$ \cite{IZ} just amounts to the
duality-invariant model recently discussed, in parallel with the BI
theory, in \cite{BN,CKR}. In our method this model corresponds to
the auxiliary algebraic equation of the fifth order. We discuss the
Legendre-type transformation of the scalar variable $\nu=V^2$ to the
$\mu$ representation \cite{IZ2}, which simplifies both solving the
auxiliary equations and constructing the Lagrangians of the
duality-invariant systems in the standard representation through the
Maxwell field strengths. We also consider the combined $(F,V,\mu)$ representation of these
Lagrangians, with both the bispinor and the scalar auxiliary fields.
The $\mu$ or $V$ representations arise after elimination of the
appropriate auxiliary field.

The authors of \cite{BN,CKO} studied  the duality-invariant theories
with higher derivatives. In Section 3 we discuss how to treat such
theories in the framework of our auxiliary-field formulation. Once
again, the basic requirement is the $O(2)$ invariance of the
interaction which is now constructed out of the auxiliary fields and
their derivatives. Though the equation for the auxiliary fields
contains derivatives, it still can be solved recursively, giving
rise to various Lagrangians with higher derivatives of the
field-strengths $F_{\alpha\beta}, F_{\dot\alpha\dot\beta}$, such
that the corresponding theories are guaranteed to exhibit duality
invariance, with the relevant generalized NGZ condition being
satisfied.

In Appendix A we present the precise correspondence between the
vector notation used in \cite{BN,CKR,CKO} and our complex
spinor notation.

In Appendix B we show that the Schr\"odinger nonlinear
constraints \cite{Schr} (pertinent to the Born-Infeld theory as a particular case of duality-symmetric  systems)
are readily reproduced in our approach.

The Lorentz non-covariant but manifestly $O(2)$ duality invariant
version of our Lagrangian with auxiliary fields is discussed in
Appendix C.

In the accompanying paper \cite{UN} we will analyze, in more
detail compared to \cite{IZ1}, the nonabelian $U(N)$ duality within
our tensorial auxiliary field formulation. In the near future we also plan to extend this formulation by incorporating scalar coset
fields which are necessary ingredients of the duality-symmetric supergravities.

\setcounter{equation}0
\section{Auxiliary variables in nonlinear electrodynamics}
\subsection{The standard duality setting}

We use the $SL(2,C)$ spinor representation for the electromagnetic
gauge field $A_{\alpha\dot\beta}$ and the complex dimensionless
Maxwell field strengths $F_{\alpha\beta}$ and
$\bar{F}_{\dot\alpha\dot\beta}$. The general nonlinear Lagrangian
with no derivatives on the field strengths is constructed from the
scalar quantities $\varphi=F^{\alpha\beta}F_{\alpha\beta}=(FF)$ and
$\bar\varphi= (\bar{F}\bar{F})$
\be
L(\varphi,\bar\varphi)=-\frac12(\varphi+\bar\varphi)+L^{int}(\varphi,\bar\varphi)\,.\lb{BaseLagr}
\ee
The physical rescaling of this dimensionless Lagrangian includes
the coupling constant $f$ of dimension $[f]=-2$ and the field
strengths $F, \bar{F}$ of the standard dimension
\be
L(F^2,\bar{F}^2)\rightarrow \frac1{f^2}L(f^2F^2,f^2\bar{F}^2)\,.
\ee

In the more familiar tensor notation (see Appendix A) the Lagrangian
$L(\varphi,\bar\varphi)$ is rewritten in terms of the standard
Maxwell field strength $F_{mn}$ and its dual $\tilde{F}_{mn}$ as
$L(\varphi,\bar\varphi)= \tilde{L}(t,z)$, where $\varphi= t+iz$ and
$t = \frac{1}{4}F_{mn}F^{mn}\,,$  $z =
\frac{1}{4}\tilde{F}_{mn}F^{mn}\,$.

The nonlinear equations of motion following from \p{BaseLagr} and
the Bianchi identities for the field strengths have the form
\bea
&&
E_{\alpha\dot\alpha}:=\partial_\alpha^{\dot\beta}
\bar{P}_{\dot\alpha\dot\beta}(F)
-\partial^\beta_{\dot\alpha} P_{\alpha\beta}(F)= 0~,\lb{NLeq}\\
&& B_{\alpha\dot\alpha} :=\partial_\alpha^{\dot\beta}
\bar{F}_{\dot\alpha\dot\beta} -\partial^\beta_{\dot\alpha}
F_{\alpha\beta}= 0~.\lb{Bian}
\eea
Here, the nonlinear bispinor fields
\bea
P_{\alpha\beta}=i\frac{\partial L}{\partial
F^{\alpha\beta}},\qquad
\bar{P}_{\dot\alpha\dot\beta}=\overline{P_{\alpha\beta}} =
-i\frac{\partial L}{\partial
\bar{F}^{\dot\alpha\dot\beta}}\lb{PbarP}
\eea
correspond to the
standard self-dual and anti-self-dual fields $G^\pm_{mn}$ used in
the tensor notation (see \p{A7}).

The set of equations \p{NLeq} and \p{Bian} is covariant under the
$O(2)$ duality transformations
\bea
\delta_\omega
F_{\alpha\beta}=\omega P_{\alpha\beta}\,,\qquad \delta_\omega
P_{\alpha\beta} =-\omega F_{\alpha\beta}\,, \lb{so2transf}
\eea
provided that the nonlinear $O(2)$ duality NGZ condition \cite{GZ2,GR} for the
Lagrangian $L$ is satisfied
\bea
F_{\alpha\beta}F^{\alpha\beta} +
P_{\alpha\beta}P^{\alpha\beta} - \mbox{c.c} = 0 \; \quad
\Longleftrightarrow \quad \varphi - \bar\varphi -
4\,[\varphi(L_\varphi)^2 -
\bar\varphi(L_{\bar\varphi})^2]=0\,.\lb{SOdual}
\eea
This condition (which is $O(2)$ invariant by itself) guarantees the consistency of
the $O(2)$ transformations \p{so2transf} with the definition
\p{PbarP}.

In the on-shell equations \p{NLeq}, \p{Bian} and the $O(2)$
transformations \p{so2transf}, the field strength $F_{\alpha\beta}$
and its conjugate are treated as independent complex variables.
 Off shell the field strengths are expressed in terms of the gauge potential $A_m\,$.
 It is impossible to implement  the $O(2)$ transformations off shell on the single real vector
field $A_m\,$, which implies that the total off-shell action cannot
possess $O(2)$ duality invariance.

Still for the Lagrangians of the duality-invariant systems there
exists the partially $O(2)$ invariant Gaillard-Zumino (GZ) representation \cite{GZ2}
\bea
L=\frac{i}2(\bar{P}\bar{F}-PF)+I(\varphi,\bar\varphi)=\frac14\tilde{G}^{mn}F_{mn}
+\tilde{I}(t,z)\,.\lb{GZ}
\eea
Here $I$ (or $\tilde{I}$) are some
$O(2)$ invariants. This representation can be easily proved by
varying the general Lagrangian \p{BaseLagr} with respect to the
transformations \p{so2transf} with taking into account the NGZ
condition \p{SOdual}, which gives
\be
\delta L = i\omega (\varphi -
\bar\varphi)\,.
\ee
This coincides with the variation of the bilinear
term in \p{GZ}; i.e. the remaining term $I$ is indeed duality-invariant.

The representation \p{GZ} on its own is not too useful since no
systematic way to construct the invariants  $I$ or $\tilde{I}$ was
known\footnote{This problem actually amounts to solving the
nonlinear NGZ condition \p{SOdual}.}. Such a way is suggested by the
formulation with the auxiliary tensorial fields.

\subsection{Reformulation through tensorial auxiliary fields}

The basic $(F,V)$-representation of the Lagrangian of general
duality-invariant system in our formalism \cite{IZ,IZ2} involves the
bispinor field-strengthes $F_{\alpha\beta}$ and
$\bar{F}_{\dot\alpha\dot\beta}$ and the auxiliary fields
$V_{\alpha\beta}$ and $\bar{V}_{\dot\alpha\dot\beta}$
\bea
&&{\cal L}(V,F)={\cal L}_2(V,F)+E(\nu,\bar{\nu})\,,\lb{LVF}\\
&&{\cal L}_2(V,F)=\frac12(F^2+\bar{F}^2)+V^2+ \bar{V}^2 - 2\,(V\cdot
F+\bar{V}\cdot\bar{F})\,.\lb{L2VF}
\eea
Here ${\cal L}_2(V,F)$ is
the bilinear part through which the Maxwell field strength
enters the action and $E(\nu, \bar\nu)$ is the nonlinear interaction
involving only auxiliary fields. We use the convenient scalar
variables
$$
\nu=V^2,  \qquad \bar\nu=\bar{V}^2\,.
$$

The $O(2)$ transformations of the involved fields read
\bea
&&\delta_\omega V_{\alpha\beta}=-i\omega V_{\alpha\beta}\,, \quad
\delta_\omega \bar{V}_{\dot\alpha\dot\beta}=i\omega
\bar{V}_{\dot\alpha\dot\beta}\,,
\quad \delta_\omega\nu=-2i\omega\nu\,, \lb{deltaV}\\
&&\delta_\omega
F_{\alpha\beta}=i\omega(F_{\alpha\beta}-2V_{\alpha\beta})\,,\quad
\delta_\omega
\bar{F}_{\dot\alpha\dot\beta}=-i\omega(\bar{F}_{\dot\alpha\dot\beta}
-2\bar{V}_{\dot\alpha\dot\beta})\,. \lb{deltaF}
\eea
The Bianchi
equation \p{Bian} together with the dynamical equation of motion
corresponding to the Lagrangian \p{LVF},
\bea
&&\partial_\alpha^{\dot\beta} \bar{P}_{\dot\alpha\dot\beta}(V,F)
-\partial^\beta_{\dot\alpha} P_{\alpha\beta}(V,F)= 0\,,\lb{eqmotVF} \\
&&P_{\alpha\beta}(F,V)= i\frac{\partial {\cal L}(V, F)}{\partial
F^{\alpha\beta}} = i(F_{\alpha\beta}-2V_{\alpha\beta})\,, \quad
\bar{P}_{\dot\alpha\dot\beta}(F,V) =
\overline{(P_{\alpha\beta}(F,V))}\,,\lb{PVF}
\eea
constitute the set
of equations manifestly covariant under these $O(2)$
transformations. The latter take the more familiar equivalent form
 just in terms of the variables $P_{\alpha\beta}, F_{\alpha\beta}$ (c.f. \p{so2transf}):
\be
\delta_\omega
P_{\alpha\beta}(F,V) = -\omega F_{\alpha\beta}\,, \quad \delta_\omega
F_{\alpha\beta} = \omega P_{\alpha\beta}(F,V)\,.
\ee

The full set of the equations of motion associated with the
Lagrangian \p{L2VF}, including the algebraic ones for the auxiliary
fields $V_{\alpha\beta}, \bar V_{\dot\alpha\dot\beta}$,  is duality-covariant
only under a certain restriction on the interaction part
$E(\nu, \bar\nu)$. This restriction actually amounts to the
constraint \p{SOdual}, but looks much simpler and has the obvious
group-theoretical meaning.

The algebraic equations for $V_{\alpha\beta}, \bar
V_{\dot\alpha\dot\beta}$ are the basic equations of our formalism.
They arise from varying ${\cal L}(V, F)$ with respect to
$V_{\alpha\beta}\,, \bar V_{\dot\alpha\dot\beta}$ and have the form
\bea
F_{\alpha\beta}=V_{\alpha\beta}+\frac12\frac{\partial
E}{\partial V^{\alpha\beta}} =V_{\alpha\beta}(1+E_\nu) \qquad
\mbox{and c.c.}\,. \lb{FVequ}
\eea
Their important corollaries are,
in particular,
\bea (\mbox{a}) \;\, V\cdot F = \nu (1 + E_\nu)\,,
\qquad (\mbox{b}) \;\,\varphi=\nu(1+E_\nu)^2\,.\lb{FVequ2}
\eea
The perturbative solution of
\p{FVequ} for functions $G$ or $L_\varphi\,$,
\bea
&&V_{\alpha\beta}(F)=F_{\alpha\beta}G(\varphi,\bar\varphi)\,,\lb{VFeq}\\
&&G(\varphi,\bar\varphi)=\frac12-L_\varphi=(1+E_\nu)^{-1}\,, \lb{EL}
\eea
allows us to construct the nonlinear Lagrangian
$L(\varphi,\bar\varphi)$ from the Lagrangian ${\cal L}(F,V(F))\,$.
Using the relations (\ref{FVequ2}b) and \p{EL}, one can also express
$\nu$ and $E_\nu$ through $\varphi$ and $L_{\varphi}$:
\be
E_\nu =
\frac{1 + 2L_\varphi}{1 - 2L_\varphi}\,, \qquad \nu  = \frac 14
\varphi(1 - 2L_\varphi)^2\,.\lb{enu}
\ee

In this way, {\it any} $L(\varphi,\bar\varphi)$ (non-singular in the
weak-field limit) can be restored by the appropriate function
$E(\nu, \bar\nu)$. The duality-invariant systems correspond to the
{\it special} choice of $E(\nu, \bar\nu)\,$. To find out the
appropriate restriction on this function, we need to rewrite the NGZ
condition \p{SOdual} in our formalism.

Using the definition \p{PVF} together with the relations \p{FVequ2}
and \p{enu}, it is easy to find
\be
F^2 + P^2 = \varphi(1 -
4L_\varphi^2) = 4 \nu E_\nu\,.
\ee
Then the nonlinear NGZ condition
\p{SOdual} is reduced to the {\it linear} constraint:
\be
F^2 + P^2
- \bar{F}^2 - \bar{P}^2 = 0\, \quad \Longleftrightarrow \quad \nu E_\nu
- \bar\nu E_{\bar\nu} = 0\,.\lb{SOdualFV}
\ee
This is none other than the condition of invariance of the
interaction $E(\nu, \bar\nu)$ under the $O(2)$ transformations
\p{deltaV}:
\bea
\delta_\omega E=2i\omega(\bar\nu E_{\bar\nu}-\nu
E_\nu)=0\,. \lb{invE}
\eea
The general solution of this condition is
obviously
\bea
E(\nu, \bar\nu) = {\cal E}(a)\,,  \quad
a=\nu\bar\nu=V^2\bar{V}^2\,. \lb{solutE}
\eea
Now, taking into account
\p{SOdualFV}, one can check the $O(2)$ covariance of the basic
 algebraic equation \p{FVequ}
for this particular subclass of the self-interactions $E(\nu,
\bar\nu)\,$. In fact, one can {\it reverse} the argument and {\it directly
deduce} \p{SOdualFV} as the necessary and sufficient condition of
covariance of Eq. \p{FVequ} under the transformations \p{deltaV},
\p{deltaF}. For what follows, it will be instructive to rewrite
\p{FVequ} and its conjugate  in the form specialized to the
duality-covariant case
\bea
F_{\alpha\beta}-
V_{\alpha\beta}=V_{\alpha\beta}\,\bar{V}^2\,{\cal E}_a(a)\,, \quad
\bar F_{\dot\alpha\dot\beta}- \bar V_{\dot\alpha\dot\beta}=\bar
V_{\dot\alpha\dot\beta}\,{V}^2\,{\cal E}_a(a)\,. \lb{FVequDu}
\eea

It is now straightforward to reveal how the general GZ
representation \p{GZ} for the Lagrangians of the duality-invariant
systems looks in our approach. We rewrite the off-shell action
\p{LVF} in the form
\bea
{\cal L}(F,V)=\frac{i}2[\bar{P}(F,V)\bar{F}-P(F,V)F]
+[V^2-(FV)]+[\bar{V}^2-(\bar{F}\bar{V})] +{\cal
E}(\nu\bar\nu)\lb{GZFV}
\eea
and observe that the $(V, F)$ analog of
the $O(2)$ invariant function $I(\varphi, \bar\varphi)$ defined in
\p{GZ} is given by
\be
I(V, F) = [V^2-(FV)]+[\bar{V}^2-(\bar{F}\bar{V})] +{\cal
E}(\nu\bar\nu)\,.
\ee
It is manifestly invariant under the duality
rotations \p{deltaV}, \p{deltaF}. Expressing $V_{\alpha\beta},
\bar{V}_{\dot\alpha\dot\beta}$ in it through $F_{\alpha\beta},
\bar{F}_{\dot\alpha\dot\beta}$ by Eqs. \p{FVequ}, we obtain the
general representation for $I(\varphi, \bar\varphi)$:
\be
I(\varphi,
\bar\varphi) = -2 a\, {\cal E}_a(a) +{\cal E}(a)\,, \qquad a =
\nu(\varphi,\bar\varphi)\bar\nu(\bar\varphi, \varphi)\,.
\ee

The basic (and most difficult) problem of the approach discussed
consists in expressing, for the given ${\cal E}(a)$, the variable
$a$ through the original variables $\varphi$ and $\bar\varphi$.
{}From the auxiliary field equation (\ref{FVequ2}a) (and its
conjugate) one can derive the real algebraic equation \cite{IZ2}
\bea (1+a{\cal
E}_a^2)^2\varphi\bar\varphi=a[(\varphi+\bar\varphi){\cal E}_a
+(1-a{\cal E}_a^2)^2]^2\,, \lb{Ebas}
\eea
which can be used for this
purpose. The general expression for the Lagrangian \p{LVF} in terms
of $\varphi$ and $\bar\varphi$, after the repeated use of the
equations \p{FVequ} and \p{FVequ2}, is obtained as follows
\bea
 L^{sd}(\varphi, \bar\varphi) = {\cal L}(V(F), F) =
 -\frac{1}{2}\,\frac{(\varphi + \bar\varphi)
 (1 - a{\cal E}_a^2) + 8a^2{\cal E}_a^3}{1 + a{\cal E}_a^2}
 + {\cal E}(a)\,, \lb{VFF}
\eea
where $a$ is related to $\varphi, \bar\varphi$ by Eq. \p{Ebas}.

To summarize,  {\it all} $O(2)$ duality-symmetric systems of
nonlinear electrodynamics without derivatives on the field strengths
are parametrized by the $O(2)$ invariant off-shell interaction of
the auxiliary fields ${\cal E}(a)$ which is a  function of the real
quartic combination of the auxiliary fields. This universality seems
to be the basic advantage of the approach with tensorial auxiliary
fields. The problem of constructing $O(2)$ duality-symmetric systems
is reduced to choosing (at will) one or another specific  ${\cal
E}(a)\,$.

After passing to the tensorial notation (see Appendix A), our basic
auxiliary field equation \p{FVequDu} (proposed more than ten years ago) is
surprisingly recognized as the ``nonlinear twisted self-duality
constraint'' of Refs. \cite{BN,CKR,CKO}, while the $O(2)$ invariant
function ${\cal E}(\nu \bar\nu)$ as the ``duality-invariant source
of deformation''. It should be pointed out that in our approach this
constraint is by no means postulated but naturally arises as the
algebraic equations of motion associated with the new off-shell
universal Lagrangian \p{LVF} for the duality-symmetric systems, in
which $E(\nu, \bar\nu) = {\cal E}(\nu\bar\nu)\,$. The auxiliary
tensorial (or bispinor) fields $V_{\alpha\beta}, \bar
V_{\dot\alpha\dot\beta}$ appearing in \p{LVF} are entirely
unconstrained off shell: there is no need to express them, e.g.,
through the second gauge field or to subject them to any other
conditions. The final nonlinear self-dual Lagrangian as a function
of the Maxwell field strengths $F_{\alpha\beta}, \bar
F_{\dot\alpha\dot\beta}$ comes out as the result of eliminating the
tensorial auxiliary fields by their equations of motion.

It is worthwhile here to recall that in \cite{GZ2,GR,HKS,BC} there
was developed another approach to solving the NGZ constraint
\p{SOdual} and restoring the appropriate Lagrangian $L(\varphi,
\bar\varphi)$, based on reducing \p{SOdual} to the  Courant-Gilbert
nonlinear differential equation. It was found that the whole family
of the duality-invariant Lagrangians is parametrized by some real
function of one argument. Despite this formal resemblance, there are
essential distinctions between this construction and our procedure.
In contrast to the former, our approach guarantees the analyticity
of $L(\varphi,\bar\varphi)$
\footnote{The perturbative linear relation of the general self-dual Lagrangian
with the corresponding real function of $(\varphi\bar\varphi)$
was analyzed in \cite{KT}.
}.
Also, at all steps we make use of the
algebraic equations, while the approach of \cite{GZ2,GR,HKS,BC}
exploits the differential equations.

Finally, we notice that, besides the continuous $O(2)$ duality, some
models of nonlinear electrodynamics reveal the so called ``discrete
duality'' or ``duality by Legendre transformation'' \cite{GZ2}. It also
has a simple realization in terms of the auxiliary fields \cite{IZ1}:
\bea
V\rightarrow -iV,\quad \nu\rightarrow -\nu,\quad
\bar{V}\rightarrow i\bar{V},\quad \bar\nu\rightarrow -\bar\nu\,,
\quad
E(\nu,\bar\nu)=E(-\nu,-\bar\nu)\,.\lb{discr}
\eea

\subsection{Alternative auxiliary field representations}
The representation \p{VFF} together with the algebraic equation
\p{Ebas} in principle solve the problem of finding the explicit form
of the action of $O(2)$ duality invariant system by a fixed ${\cal
E}(a)$ in terms of the Maxwell strengths  (at least, as a power
series in $\varphi$ and $\bar\varphi\,$). However,  in practice it
is frequently more convenient to deal with the set of the auxiliary
complex variables $\mu$ and $\bar\mu$ instead of the original ones
$\nu$ and $\bar\nu$. These two sets of variables are related to each
other through the Legendre transformation
\bea
&& \mu :=E_\nu,\quad
\bar\mu := E_{\bar\nu}\,,\qquad  E-\nu E_\nu-\bar\nu E_{\bar\nu} :=
H(\mu,\bar\mu), \lb{mudef} \\
&& \nu=-H_\mu,\quad \bar\nu=-H_{\bar\mu}\, \qquad E = H -\mu H_\mu -
\bar\mu H_{\bar\mu}\,. \lb{EH}
\eea

The function  $H(\mu, \bar\mu)$ provides an alternative
representation of the nontrivial interaction $E\neq 0$. Under the
off-shell $O(2)$ duality the variables $\mu$ and $\bar\mu$ are
evidently transformed as
\be
\delta_\omega \mu = 2i\omega \mu\,,
\qquad \delta_\omega \bar\mu = -2i\omega \bar\mu\,,\lb{mutrans}
\ee
so the $O(2)$ duality invariance requires
$$
H(\mu, \bar\mu) = I(b)\,, \quad  b = \mu\bar\mu\,.
$$
{}From \p{mudef}, \p{EH} one finds
\bea
{\cal E}(a)-2a{\cal E}_a(a)
=I(b)\,, \quad a=bI_b^2, \quad b = a{\cal E}_a^2\,, \qquad {\cal
E}_{a}=-(I_b)^{-1}. \lb{relab}
\eea
In order to guarantee the $(\mu, \bar\mu)
\leftrightarrow (\nu, \bar\nu)$ transform to be invertible, we are
led to assume that ${\cal E}_a(a=0) \neq 0\,, \; I_b(b=0) \neq 0$,
i.e. that ${\cal E}(a)$ starts with the term $\propto a\,$.

The basic algebraic equations of this representation directly follow
from Eq. (\ref{FVequ2}b)
\be \varphi=-(1+\mu)^2\,H_\mu=-(\bar\mu
+2b+ b\,\mu)\, I_b \quad \mbox{and c.c.}\,.\lb{vpmurel}
\ee
They enable expressing $\mu, \bar\mu$ in terms of $\varphi,
\bar\varphi\,$.  The equation for the invariant variable $b$ also
directly  follows from \p{Ebas}, and it reads \cite{IZ2}
\bea
(b+1)^2\,\varphi\bar\varphi=b\,[\varphi+\bar\varphi-I_b(b-1)^2]^2.\lb{bequ}
\eea
The relevant expression for the general Lagrangian ${\cal L}(V(F), F)$ can
be easily obtained from
the expression \p{VFF}:
\bea
L^{sd}(\varphi, \bar\varphi) = -\frac{1}{2}\,(\varphi + \bar\varphi
+ 4bI_b)\,\frac{1-b}{1+b} + I(b)\,.\lb{bsd}
\eea

It is worth noting that one can reproduce the algebraic relations
\p{vpmurel} from the new Lagrangian with $\mu$ as an independent
complex auxiliary field:
\bea
&&\tilde{L}(\varphi,\mu)=\frac{\varphi(\mu-1)}{2(\mu +1)}
+\frac{\bar\varphi(\bar\mu-1)}{2(\bar\mu +1)}+I(\mu\bar\mu).\lb{Lvm}
\eea
Varying \p{Lvm} with respect to $\mu, \bar\mu$ yields just
\p{vpmurel}, from which one can derive Eq. \p{bequ} without any
reference to the original $\nu, \bar\nu$ representation. Eliminating
$\mu$ and $\bar\mu$ from \p{Lvm} in terms of $\varphi, \bar\varphi$
and $b$ with the help of \p{vpmurel} and its corollary \p{bequ}, we
recover the Lagrangian \p{bsd}.

The dual field strength $P_{\alpha\beta}$ appearing in the nonlinear
equation of motion for $F_{\alpha\beta}$ derived directly from the
extended Lagrangian \p{Lvm} is as follows
\bea
&&P_{\alpha\beta}(F,\mu) =2iF_{\alpha\beta}\frac{\partial\tilde{L}(\varphi,\mu)}{\partial\varphi}
=iF_{\alpha\beta}\,
\frac{\mu-1}{\mu +1}\,.
\eea
The consistency of the nonlinear $O(2)$ transformations,
\be
\delta_\omega F_{\alpha\beta} =\omega
P_{\alpha\beta}\,,\qquad\delta_\omega P_{\alpha\beta}=-\omega
F_{\alpha\beta}\,,
\ee
can be easily checked, assuming that the auxiliary variables
$\mu, \bar\mu$ transform as in \p{mutrans}.

The Lagrangian \p{Lvm} provides the convenient  off-shell
description of the general duality-invariant system of nonlinear
electrodynamics with an extra complex auxiliary field. The free
Lagrangian corresponds to the limit $\mu=0\,$,
$\tilde{L}(\varphi,0)=-\frac12(\varphi+\bar\varphi)\,$.

There also exists a combined representation for  the general
self-dual Lagrangian, with two auxiliary fields, $V_{\alpha\beta}$
and $\mu\,$,
\bea
{\cal L}(F,V,\mu)=
\frac{1}2(F^2+\bar{F}^2)-2(VF)-(\bar{V}\bar{F})+V^2(1+\mu)+(1+\bar\mu)\bar{V}^2
+I(\mu\bar\mu).
\eea
Eliminating the bispinor auxiliary field via its algebraic
equation $V_{\alpha\beta}=F_{\alpha\beta}/(1+\mu)\,$, we come back
to the Lagrangian \p{Lvm}. On the other hand, the $(\mu, \bar\mu)$
equations give us just the relations \p{EH} between $\nu=V^2\,,
\bar\nu=\bar{V}^2$ and $\mu\,, \bar\mu$
\bea
\nu+\bar\mu
I_b=0\,,\qquad \bar{\nu}+\mu I_b=0\,.
\eea
The equations for $\mu(\nu,\bar\nu)$ can be solved recursively in the general case
and explicitly for some special functions $I_b$, thus yielding the
original $(F,V)$ representation.

\subsection{Examples}
\noindent{\it I. Born-Infeld model}. The classical example of the
$O(2)$ duality invariant model of nonlinear electrodynamics is the
renowned Born-Infeld theory. In our approach it has a more simple
description in the $\mu$ (or $b$) representation. The corresponding
function $I(b)$ is as follows
\bea
I^{BI}(b)=\frac{2b}{b-1}\,,\qquad
I^{BI}_b=-\frac{2}{(b-1)^2}\,. \lb{BII}
\eea
The equation \p{bequ}
becomes quadratic in this case:
\bea
\varphi\bar\varphi\,
b^2+[2\varphi\bar\varphi-(\varphi+\bar\varphi+2)^2]\,b+
\varphi\bar\varphi=0\,.
\eea
It can be explicitly solved as
\cite{IZ2}:
\bea
b = \frac{4\varphi\bar\varphi}{[2(1 + Q) +\varphi +
\bar\varphi]^2}\,, \quad Q(\varphi) = \sqrt{1 + \varphi +
\bar\varphi + (1/4)(\varphi - \bar\varphi)^2}\,. \lb{bBI}
\eea
After substituting \p{BII}, \p{bBI} into \p{bsd} we recover the standard
Born-Infeld Lagrangian
\be
L^{BI} (\varphi, \bar\varphi) = 1 -
\sqrt{1 + \varphi + \bar\varphi + (1/4)(\varphi -
\bar\varphi)^2}\,.\lb{LagrBI}
\ee

The formulation of the BI theory in the original $a$ (or $\nu$)
representation is also possible but it proves to be much more
involved. The original variable $a$ is related to $b$ as
$$
a = \frac{4b}{(1-b)^4}
$$
and for ${\cal E}(a)$ one obtains the representation
\be
{\cal E}^{BI}(a) = \frac{2b(a)[1+b(a)]}{[1-b(a)]^2} = 2[2t^2(a) + 3t(a)
+1]\,, \lb{Ebi}
\ee
where $t(a)$ is defined by the 4-th order equation
$$
t^4 + t^3 -(1/4)a =0, \quad t(\nu = \varphi =0) = -1\,.
$$
Solving it, one can find closed expressions for both $t(a)$ and
${\cal E}^{BI}(a)$, but they look not too illuminating.
Up to the 3d order in $a$:
\be
{\cal E}^{BI}(a) = \frac12\,a - \frac18\,a^2 +
\frac{3}{32}\,a^3 + O(a^4)\,. \lb{3BI}
\ee

\vspace{0.3cm}

\noindent{\it II. The simplest interaction (SI) model}. This $O(2)$ duality invariant
model is one of those considered in \cite{BN} and referred to as the
BN model in \cite{CKR}. It corresponds to the following constraint
on the Lagrangian $L^{SI}(\varphi, \bar\varphi) =
{\tilde{L}}^{BN}(t, z)$ \cite{CKR}:
\be
(1+\tilde{L}{}^{SI}_t-i\tilde{L}{}^{SI}_z)-\frac18(t-iz)(1-\tilde{L}{}^{SI}_t
-i\tilde{L}{}^{SI}_z)^2(1-\tilde{L}{}^{SI}_t+i\tilde{L}{}^{SI}_z)=0\,.
\ee

It can be rewritten through the variables $\varphi, \bar\varphi$ as
\be
(1+2L^{SI}_\varphi)-\frac18\bar\varphi(1-2L_{\bar\varphi}^{SI})^2
(1-2L^{SI}_\varphi)
= 0 \qquad \mbox{and c.c.}\,,\lb{BNconstr}
\ee
where we made use of the correspondence
$$
\partial_\varphi = \frac{1}{2}(\partial_t -i \partial_z)\,, \qquad \partial_{\bar\varphi} = \frac{1}{2}(\partial_t + i \partial_z)\,.
$$
Using the general  relations \p{enu}, as well as their corollaries
$$
1 + 2L_\varphi = \frac{2E_\nu}{1+ E_\nu}\,, \quad 1 - 2L_\varphi =
\frac{2}{1+ E_\nu}\,,
$$
it is straightforward to rewrite the constraint \p{BNconstr} in
terms of $\nu, \bar\nu$ and $E_\nu$, and to find that in the new
setting  this constraint is reduced to the pretty simple condition
\be
E_\nu = \frac{1}{2}\bar\nu \quad \Rightarrow \quad E^{SI}(\nu,
\bar\nu) = \frac12 \nu\bar\nu= \frac12\, V^2\bar{V}{}^2 = \frac12\,
a\,.
\ee
Thus the BN model amounts to the simplest choice of the
$O(2)$ invariant auxiliary interaction function ${\cal E}(a)$, the
one quartic in the auxiliary tensor fields. This model was actually analyzed
in detail in our paper \cite{IZ} as
the simplest example of the auxiliary interaction generating the
non-polynomial self-dual electromagnetic Lagrangian in the $F$
representation. The relevant interaction in the $b$-representation
is also linear,
\be I^{SI}(b) = -2 b\,, \quad I^{SI}_b = -2\,.
\ee

Despite such a simple off-shell form of the auxiliary interaction,
it is hardly possible to find a closed on-shell form of the
nonlinear Lagrangian $L^{SI}(\varphi, \bar\varphi)$, since the
algebraic equations relating $a$ and $b$ to $\varphi, \bar\varphi$
are of the 5-th order. E.g., the equation \p{bequ} becomes
\be
(b +1)^2\,\varphi\bar\varphi = b\,[\varphi + \bar\varphi +2(b-1)^2
]^2\,.
\ee
Nevertheless, it is straightforward to solve these
equations as infinite series in  $\varphi, \bar\varphi$ and then to
restore $L^{BN}(\varphi, \bar\varphi)$ to any order using the
representations \p{VFF} or \p{bsd}. In the present case it is
simpler to solve the coupled set of the algebraic equations for the
function $G(\varphi, \bar\varphi)$ defined in \p{EL}:
\bea
G=\frac1{1+\frac12\bar\nu}=\frac1{1+\frac12\bar\varphi\,\bar{G}^2}\,,
\quad \bar G = \frac1{1+\frac12\varphi\,{G}^2}\,, \lb{Gequ}
\eea
which amounts to the following 5-th order equation for $G$:
\bea
2\left(1+\frac12\varphi
G^2\right)^2=G\left[\bar\varphi+2\left(1+\frac12\varphi
G^2\right)^2\right].\lb{Gequ1}
\eea
For the variables $a$ and $b$ one gets, respectively,
$$
a = \varphi\bar\varphi\,G^2\bar G^2\,, \qquad b= \frac14\, a\,.
$$

We give the self-dual Lagrangian $L^{SI}(\varphi, \bar\varphi)$ up
to 10-th order in $F_{\alpha\beta}$ and $F_{\dot\alpha\dot\beta}$,
in parallel with the analogous expansion of $L^{BI}(\varphi,
\bar\varphi)$. To this end, we use the expansion of ${\cal E}(a)$ in
powers of $a$ up to the second order :
\bea
{\cal E}(a)=e_1a+\frac12e_2a^2
+O(a^3)\,,\quad
{\cal E}_a=e_1+e_2a
+O(a^2)\,.
\lb{expanE}
\eea
Here $e_1,~ e_2 , \ldots$ are some real
coefficients. All coefficients $e_k$ are non-vanishing for the BI
model: $e_1 = \frac12, \; e_2 = -\frac14,\ldots$ \cite{IZ}, whereas
all $e_k$ except for $e_1 = \frac12$ are vanishing for the SI model.
Knowing ${\cal E}(a)$ up to the second order in $a$ allows one to
uniquely  restore $L^{sd}(\varphi, \bar\varphi)$ up to the 10th
order in $F_{\alpha\beta}, \bar F_{\dot\alpha\dot\beta}$:
\bea
&&L^{sd}=-\frac12(\varphi+\bar\varphi) +e_1\varphi\bar\varphi
-e_1^2(\varphi^2\bar\varphi+\varphi\bar\varphi^2)
+e_1^3(\varphi^3\bar\varphi+\varphi\bar\varphi^3)+(4e_1^3+\frac12e_2)\varphi^2\bar\varphi^2
\nn
&& -e_1^4(\varphi^4\bar\varphi+\varphi\bar\varphi^4)
-(10e_1^4+2e_1e_2)(\varphi^3\bar\varphi^2+\varphi^2\bar\varphi^3)+O(F^{12}).\lb{expanL}
\eea
The Lagrangian for the BI model corresponds to the choice $e_1
= \frac12, \; e_2 = -\frac14$ in \p{expanL} (c.f. \p{LagrBI}), while
the Lagrangian of the SI (BN) model is reproduced with  $e_1 =
\frac12, \; e_2 = 0\,$.
The Lagrangians for the BI and BN models as the power series in the
tensor variables $t$ and $z$ were given in \cite{CKR}.

\vspace{0.3cm}

\noindent{\it III. More examples}. In \cite{IZ2} we also considered
the following simple ansatz for the one-parameter deformation of the
BI auxiliary Lagrangian \p{BII}
\be
I_b=-\frac{2-cb}{(b-1)^2}\,.\lb{BIdefor}
\ee
This ansatz gives rise to some 3-rd order equation for $b\,$.
The simplest choice is $c=2\,$, for which
\be
I_b=-2/(1-b)\,,\quad
I(b)=2\ln(1-b)\,.\lb{Ic2}
\ee
The corresponding $\varphi, \bar\varphi$
representation for the Lagrangian involves only one unknown function
$r=\mbox{Re}\,\mu\,$,
\bea
&& \mu =
r-\frac{(\varphi-\bar\varphi)}{4}\,, \quad
b=r^2-\frac{(\bar\varphi-\varphi)^2}{16}\,, \nn
&& L =
-(\mu+\bar\mu)+2\ln(1-b)=-2r-2\ln\left[1-r^2+\frac{(\bar\varphi-\varphi)^2}{16}\right].
\lb{Lthree}
\eea
The basic algebraic equation in the $\mu$-representation \p{bequ} is
reduced to the cubic equation for $r$:
\bea
&&r^3+(2+t)
r^2+(1+\frac1{4}z^2)r-\frac12t+\frac1{4}(2+t)z^2=0\,, \lb{requ}
\eea
where $t = (\varphi + \bar\varphi)/2\,, z = (\varphi -
\bar\varphi)/2i$ (see Eq. (A.10)). One can explicitly solve Eq.
\p{requ} using the Cardano formula or stick to its perturbative
solution.

It is easy to obtain the $(F,V)$ representation \p{LVF} for the
Lagrangian of the considered system. The corresponding $O(2)$
invariant function ${\cal E}(a)$ is given by the following
expression
\bea
&& {\cal E}(a)=2(\sqrt{1+a}-1) -2\ln\left[\frac12(1 +
\sqrt{1+a})\right]
=\frac12a-\frac1{16}a^2-\frac1{6}a^3+O(a^4)\,, \nonumber \\
&& {\cal E}_a =\frac{1}{1+\sqrt{1+a}}\,. \nonumber \eea

In \cite{IZ2} we also studied some duality-invariant systems which
require solving the 4-th order equations.

\section{Auxiliary variables in duality-invariant theories with higher derivatives}
\setcounter{equation}{0}

\subsection{Generalities}
The nonlinear electromagnetic Lagrangians with higher derivatives
are functions of the variables
\bea
F\,,\quad \partial_m F\,,\quad
\partial_m\partial_n F, \quad \partial_m\partial_n\partial_r F\ldots
\eea
and their complex conjugates. Derivatives of the field
strengths appear with some coupling constants of non-trivial
dimensions. The higher-derivative electromagnetic Lagrangians in the
explicit form involve various scalar combinations of these
variables, e.g.,
\be
F^2, \;\;(\partial^mF\partial_mF)\,, \;\;
(\partial^mF^2\partial_mF^2)\,,\quad (F\Box^N F)\,,\ldots\,.
\ee
It is known that the higher-derivative generalizations of the
duality-invariant Lagrangians contain all orders of derivatives of
$F_{\alpha\beta}$ and $\bar F_{\dot\alpha\dot\beta}$ \cite{BN,CKO}.

The nonlinear equations of motion of such generalized Lagrangians
are expressed through the Lagrange derivative \cite{KT}
\bea
&&P_{\alpha\beta}\equiv
(G^+)_{\alpha\beta}=\frac12(\Sigma^{mn})_{\alpha\beta}G^+_{mn}
=i\frac{\Delta L}{\Delta F^{\alpha\beta}}\,,\\
&&\frac{\Delta L}{\Delta {F}^{\alpha\beta}}=\frac{\partial
L}{\partial{F}^{\alpha\beta}} -\partial_m\frac{\partial
L}{\partial(\partial_m {F}^{\alpha\beta})}
+\partial_m\partial_n\frac{\partial L}{\partial(\partial_m\partial_n
{F}^{\alpha\beta})} +\ldots\,.
\eea

The definition of the $O(2)$ duality transformations and finding out
the appropriate Lagrangians which would yield the duality-symmetric
systems of equations with higher derivatives is a rather difficult
task (see, e.g.,\cite{BN,CKO}). We propose to tackle this problem
in the framework of the higher-derivative generalization of our
formulation with the auxiliary bispinor fields.

This generalization is accomplished rather straightforwardly: we start from the
original $(F, V)$ Lagrangian \p{LVF} with $O(2)$ invariant
self-interaction ${\cal E}(\nu\bar\nu)$ and allow the latter to
depend also on the derivatives of auxiliary fields (still keeping
invariance under the $O(2)$ transformations \p{deltaV}):
\bea
&&{\cal
L}(F,V,\partial V, \partial^2 V, \ldots)={\cal L}_2+ {\cal
E}(V,\partial V, \partial^2 V, \ldots)\,.\lb{FVH}
\eea
Here ${\cal L}_2$ is the old bilinear part of the Lagrangian \p{L2VF}.
By construction, this modified Lagrangian admits an analog of the GZ
representation \p{GZFV}
\bea
{\cal L}(F,V,\ldots)=\frac{i}2[\bar{P}(F,V)\bar{F}-P(F,V)F] +V^2-(FV)+
\bar{V}^2-(\bar{F}\bar{V}) +{\cal E}(V,\ldots)\lb{GZFVH}
\eea
for any ${\cal E}$. The equations of motion for this Lagrangian
contain the Lagrange derivative of ${\cal E}$
\bea
&&\partial^\alpha_{\dot\beta}(F-2V)_{\alpha\beta}+\partial^{\dot\alpha}_{\beta}
(\bar{F}-2\bar{V})_{\dot\alpha\dot\beta} = 0\,, \lb{FVHeq1} \\
&&F_{\alpha\beta}=V_{\alpha\beta}+\frac12\frac{\Delta {\cal E}}
{\Delta V^{\alpha\beta}}\,.\lb{FVHeq}
\eea
It is easy to see that this
set of  equations together with the Bianchi identity \p{Bian} is
covariant under the $O(2)$ duality transformations \p{deltaF},
\p{deltaV}, provided that ${\cal E}(V, \partial V)$ is $O(2)$
invariant,
\be
\delta_\omega\, {\cal E}(V, \partial V, \partial^2 V,
\ldots) = 0\,.
\ee

In the simplest non-trivial case, when the auxiliary interaction
depends only on $V, \bar V$ and their first derivatives, i.e. ${\cal
E} = {\cal E}(V, \partial V)\,$, we have
\bea
&&\frac{\Delta {\cal
E}}{\Delta V^{\alpha\beta}}=\frac{\partial {\cal E}}{\partial
V^{\alpha\beta}} -\partial_m\frac{\partial {\cal
E}}{\partial(\partial_m V^{\alpha\beta})}\,.
\eea
Using the definitions
\be
P_{\alpha\beta}(F,V)=i(F-2V)_{\alpha\beta}\,,\quad
\bar{P}_{\dot\alpha\dot\beta}(F,V)=-i(\bar{F}-2\bar{V})_{\dot\alpha\dot\beta}
\ee
and Eq.\p{FVHeq}, we obtain the relations
\bea
&&P^2+F^2=4(FV)-4V^2=2V^{\alpha\beta}\frac{\Delta {\cal E}}{\Delta
V^{\alpha\beta}}\nn
&&=\,2V^{\alpha\beta}\frac{\partial {\cal
E}}{\partial V^{\alpha\beta}}
+2\partial_mV^{\alpha\beta}\frac{\partial {\cal
E}}{\partial(\partial_m V^{\alpha\beta})}
-2\partial_m\left[V^{\alpha\beta}\frac{\partial {\cal E}}
{\partial(\partial_m V^{\alpha\beta})}\right].
\eea
Now we can prove
the integral form of the NGZ identity, using the $O(2)$ invariance
of the interaction ${\cal E}(V, \partial V)$
\bea
&&\int
d^4x[P^2(F,V)+F^2-\bar{P}^2(F,V)-\bar{F}^2]=\int d^4x[\delta_\omega
{\cal E}+\mbox{div}]=0\,.
\eea
Substituting the solution of Eq.\p{FVHeq} into this relation, we obtain the nonlinear NGZ
identity for the original $F$-representation with higher
derivatives.

The same NGZ identity is valid in the general case, for an arbitrary
$O(2)$ invariant auxiliary self-interaction ${\cal E}(V, \partial V,
\partial^2 V, \ldots)\,$.

Now we turn to the examples.

\subsection{The simplest case}
The simplest real invariant with two derivatives is
\bea
&&{\cal
E}(V, \partial V) = \rho=\bar\rho \sim
\partial_\beta^{\dot\beta}V^{\alpha\beta}\partial^{\dot\xi}_{\alpha}
\bar{V}_{\dot\beta\dot\xi}\,,\quad \delta_\omega \rho=0\,.
\eea
In the tensor representation the same $O(2)$ invariant reads as  $\rho\sim
\partial^nV^+_{mn}\partial_rV^{-rm}$. The relevant equations for $V_{\alpha\beta}$
and its conjugate are
\bea
F_{\alpha\beta}=V_{\alpha\beta}
+\frac{1}{2}b_1\partial_\beta^{\dot\beta}\partial^{\dot\rho}_{\alpha}
\bar{V}_{\dot\beta\dot\rho}\,, \quad
\bar{F}_{\dot\alpha\dot\beta}=\bar{V}_{\dot\alpha\dot\beta}
+\frac{1}{2}b_1\partial_{\dot\beta}^{\beta}\partial^{\rho}_{\dot\alpha}
{V}_{\beta\rho}\,,
\eea
where $b_1$ is a real constant. These
equations have the non-local solution
\bea
V_{\alpha\beta}=\frac{1}{1-\frac14b^2_1\Box^2}\left[F_{\alpha\beta}-\frac12b_1
\partial_\beta^{\dot\beta}\partial^{\dot\rho}_{\alpha}
\bar{F}_{\dot\beta\dot\rho}\right].
\eea
The corresponding $L(\varphi,
\bar\varphi)$ is an extension of the standard Maxwell Lagrangian by
nonlocal higher-derivative terms of the second order in the field
strengths $F_{\alpha\beta}, \bar{F}_{\dot\alpha\dot\beta}\,$. Although
such nonlocal modifications of the free bilinear action could appear
as duality-invariant counterterms in extended supergravities
\cite{BN}, in what follows we assume that the higher-derivative
terms appear only at the interaction level, i.e. that the bilinear
part of the full action still coincides with the standard Maxwell
action. Under this natural assumption, ${\cal E}(V, \partial V,
\partial^2 V, \ldots)$ contains no terms bilinear in
$V_{\alpha\beta}, \bar{V}_{\dot\alpha\dot\beta}\,$. The relevant
auxiliary-field equations can always be solved by recursions,
yielding no non-localities in the perturbative expansions and
producing $L(\varphi, \bar\varphi)$ in which all higher-derivative
terms come out only in the interaction part.

In the remainder of this Section we will present a few specific
examples of the duality-invariant higher-derivative systems of this
type.

\subsection{Auxiliary interaction with two derivatives}
As the first example we consider the simplest quartic mixed
interaction with two derivatives
\bea
{\cal
E}_{(2)}=\frac12\nu\bar\nu+c\partial^m\nu\partial_m\bar\nu, \quad
\frac{\Delta {\cal E}_{(2)}}{\Delta
V^{\alpha\beta}}=2V_{\alpha\beta}[1+\frac12\bar\nu
-c\Box\bar\nu]\,,\lb{2quart}
\eea
where $c$ is a coupling constant of
dimension $-2$ and $\Box = \partial^m\partial_m\,$. The auxiliary
field equation has the form
\bea
&&F_{\alpha\beta}=V_{\alpha\beta}(1+\frac12\bar\nu-c\Box
\bar\nu)\,,\quad \Box\bar\nu= 2(\bar{V}^{\dot\alpha\dot\beta} \Box
\bar{V}_{\dot\alpha\dot\beta}+\partial^m\bar{V}^{\dot\alpha\dot\beta}
\partial_m \bar{V}_{\dot\alpha\dot\beta})\,.
\eea
The perturbative solution contains the higher derivatives
\bea
V^{(1)}_{\alpha\beta} &=& F_{\alpha\beta}\,, \nn
V^{(3)}_{\alpha\beta} &=&
-F_{\alpha\beta}(\frac12\bar\varphi-c\Box\bar\varphi)\,,\nn
V^{(5)}_{\alpha\beta} &=&
F_{\alpha\beta}\{\frac12\bar\varphi(\frac12\bar\varphi-c\Box\bar\varphi)
+\varphi(\frac12\bar\varphi-c\Box\bar\varphi)-c\Box \bar\varphi\,
(\frac12\bar\varphi-c\Box\bar\varphi)\nn
&&-\,\,
2c\Box[\bar\varphi(\frac12\varphi-c\Box \varphi)]\}\,, \qquad
\mbox{etc}\,.
\eea
The order of derivatives grows with each next
recursion. The self-dual actions in the $F$-representation contains
all orders of higher derivatives distributed over terms of all
orders in the Maxwell field strengths. The perturbative construction
of these actions, though being quite algorithmic, is much more
involved technically as compared to the case without derivatives.
For our model we obtain, up to terms of 8-th order,
\bea
L(F,\partial^{N}F) &=& -\frac12(\varphi+\bar\varphi)
+\frac12\varphi\bar\varphi-\frac14\varphi^2\bar\varphi-\frac14\varphi\bar\varphi^2 + c\partial^m\varphi\partial_m\bar\varphi
\nn
&&+\,c\varphi\bar\varphi(\Box\varphi
+\Box\bar\varphi)-c^2\varphi(\Box\bar\varphi)^2-c^2\bar\varphi(\Box\varphi)^2+O(F^{8})\,.
\eea

The model \p{2quart} can be regarded as the ``minimal''
higher-derivative deformation of the SI (or BN) model of Sect.3.
Similarly, we could choose, as the undeformed ${\cal E}$, the
interaction \p{Ebi} corresponding to the BI theory. Adding to it the
same term bilinear in derivatives as in \p{2quart}, we would get the
``minimal'' higher-derivative duality-invariant deformation of the
BI theory.

\subsection{Auxiliary interaction with four derivatives}

Simple examples of the duality-invariant systems with higher
derivatives were analyzed in \cite{BN,CKO}, based on the
``nonlinear twisted duality constraint'' which is equivalent to our
auxiliary field equation \p{FVHeq}.

In our language these systems correspond to the quartic $O(2)$
invariant interactions
\bea A_4= A(\partial^mV\partial^nV)
(\partial_m\bar{V}\partial_n\bar{V})\,,\qquad
B_4=B(\partial^mV\partial_mV) (\partial^n\bar{V}\partial_n\bar{V})\,,
\eea
where $A$ and $B$ are some constants of dimension $-4$ and
brackets denote traces with respect to  the $SL(2,C)$ doublet
indices.

We consider the construction of the Lagrangian in the
$F$-representation for the model $A_4$. The basic equation \p{FVHeq}
in this case is reduced to
\bea
&&F_{\alpha\beta}=V_{\alpha\beta}-A\partial^m\left[
\partial^nV_{\alpha\beta}
(\partial_m\bar{V}\cdot\partial_n\bar{V})\right].
\eea
The first two
terms in its perturbative solution are as follows
\bea
&&V^{(1)}_{\alpha\beta}=F_{\alpha\beta},\quad
V^{(3)}_{\alpha\beta}=A\partial^m\left[
\partial^nF_{\alpha\beta}
(\partial_m\bar{F}\cdot\partial_n\bar{F})\right].
\eea
The corresponding Lagrangian in the $F$-representation involves higher
derivatives, starting from the sixth order in fields
\bea
L^{(6)}=(V^{(3)}V^{(3)})-2A[(V^{(3)}\partial^n\partial^mF)(\partial_m\bar{F}\partial_n\bar{F})
+(V^{(3)}\partial^mF)\partial^n(\partial_m\bar{F}\partial_n\bar{F})]
+\mbox{c.c.}\,.
\eea

An alternative example of the auxiliary interaction with four
derivatives which was not considered in \cite{CKO} reads (just for a
change, we use here the tensor notations, see Appendix A):
\bea
&&C_4=C\,\partial^nV^+_{mn}\partial_rV^{-mr}\partial^pV^+_{qp}\partial_sV^{-qs}\,.
\eea
Here $C$ is a constant of dimension $-4\,$.

The basic auxiliary equation of this model reads
\bea
&&F^{+mn}=V^{+mn}-2C\pi^{mn}_{uv}\partial^u
\left[\partial_rV^{-vr}\partial^pV^+_{qp}
\partial_sV^{-qs}\right],
\eea
where
\bea
\pi^{mn}_{uv}=\frac14 (\delta^m_u\delta^n_v-
\delta^m_v\delta^n_u +{i}\varepsilon^{mn}_{uv})
\eea
is the  projector on the self-dual part. The equation for
$F^-_{mn}$ can be obtained by complex conjugation. The  first two
terms of the perturbative solution are given by the expressions
\bea
V_{mn}^{+ (1)}=F^{+}_{mn},\qquad V_{mn}^{+ (3)}=
2C\pi_{mn}^{uv}\partial_u\left[\partial^rF^{-}_{vr}\partial^pF^+_{qp}
\partial_sF^{-qs}\right].
\eea
The perturbative construction of the Lagrangian in the $F$-representation
is completely analogous to the previous cases.
\setcounter{equation}{0}

\section{Conclusions}

In this paper, following and extending our earlier results
\cite{IZ,IZ2}, we presented the evidence that {\it all}
duality-invariant systems of nonlinear electrodynamics (including those with
higher derivatives) admit an off-shell formulation with the
auxiliary bispinor (tensorial) fields. These fields are fully
unconstrained off shell, there is no need to express them through
any second gauge potentials, etc.

The whole information about the given duality-invariant system is
encoded in the $O(2)$ invariant interaction function which depends only on the auxiliary
fields (or also on their derivatives) and can be chosen {\it at will}. In many cases it looks much simpler compared to
the final on-shell action written in terms of the Maxwell field strengths. The renowned nonlinear NGZ constraint is {\it linearized}
in the new formulation and becomes just the requirement of $O(2)$ invariance of the auxiliary interaction.
The $O(2)$ (and in fact $U(N)$ \cite{IZ1,UN}) duality symmetry is realized off shell by {\it linear} transformations.

The basic algebraic equations eliminating the auxiliary tensor fields are equivalent to the recently employed
{\it ``nonlinear twisted self-duality constraint''} \cite{BN,CKR,CKO}. In our approach this constraint appears
as the {\it equations of motion} \p{FVequDu}, \p{FVHeq} associated with the well defined off-shell Lagrangians.
Based on the relations \p{A15} - \p{A18}, it is instructive here to give the dictionary of correspondence between
the notations used in \cite{CKR,CKO} and our bispinor notation (up to numerical coefficients)
\bea
 \left( T^+_{mn}\,, \; T^{*-}_{mn}\right) \; \Leftrightarrow  \;  \left[ (F-V)_{\alpha\beta}\,, \;(\bar F- \bar V)_{\dot\alpha\dot\beta}\right], \quad
 \left( T^{*+}_{mn}\,, \; T^{-}_{mn}\right) \; \Leftrightarrow  \;  \left( V_{\alpha\beta}\,, \; \bar V_{\dot\alpha\dot\beta}\right), \quad
{\cal I}^{(1)}\; \Leftrightarrow  \; {\cal E}\,. \nonumber
\eea

Let us finish by mentioning a few proposals for the future study.
\vspace{0.2cm}

First, it is of urgent interest to extend our formulation to the
${\cal N}=1$, ${\cal N}=2$, $\ldots$ supersymmetric duality-invariant systems,
including supersymmetric Born-Infeld theories, in both the flat and
the supergravity backgrounds.

Second, it is desirable to learn how the scalar and other fields
can  be self-consistently inscribed into the framework with the
auxiliary tensorial fields.

Finally, we would like to point out that the formulation with the
auxiliary fields suggests  a  {\it new view} of the
duality-invariant systems: it seems natural, in both the classical and the
quantum cases, {\it not} to eliminate the tensorial auxiliary fields
by their equations of motion beforehand, but to deal with the
off-shell actions {\it at all steps}. This would  as well refer to
the quantum counterterms which should appear in this approach as
corrections to the original auxiliary interaction function ${\cal E}\,$.
It is worthwhile here to recall the off-shell superfield
approach to supersymmetric theories, which in many cases radically
facilitates the quantum calculations and unveils the relevant
intrinsic geometric properties without any need to pass on shell
through eliminating the auxiliary fields. It is also worth recalling
that the tensorial  auxiliary fields have originally appeared  just
within the off-shell superfield formulation of ${\cal N}=3$
supersymmetric Born-Infeld theory \cite{IZ}.

\section*{Acknowledgements}
We acknowledge a partial support from the RFBR grants Nr.12-02-00517,
Nr.11-02-90445, the grant DFG LE 838/12-1 and a grant from
 the Heisenberg-Landau program.

\setcounter{equation}0
\renewcommand{\theequation}{A.\arabic{equation}}
\section*{A. Spinor and tensor notations in  electrodynamics}

We use the Minkowski metric $\eta_{mn}=\mbox{diag}(1, -1, -1,-1)$
and the $2\times 2$ Weyl matrices
\bea
&&(\sigma^m)_{\alpha\dot\beta}\,,\quad
(\bar\sigma^m)^{\dot\alpha\beta}=
\varepsilon^{\beta\alpha}\varepsilon^{\dot\alpha\dot\beta}(\sigma^m)_{\alpha\dot\beta}\,,
\\
&&(\sigma^m\bar\sigma^n+\sigma^n\bar\sigma^m)_\alpha^\beta=2\eta^{mn}\delta_\alpha^\beta\,,\\
&&\sigma^n\bar\sigma^s\sigma^m=\eta^{ns}\sigma^m-\eta^{nm}\sigma^s+\eta^{sm}\sigma^n
-i\varepsilon^{nsmr}\sigma_r\,.
\eea

The vectors in the spinor and tensor notations are related as
\be
A_{\alpha\dot\beta}=(\sigma^m)_{\alpha\dot\beta}A_m\,,\quad
\partial_{\alpha\dot\beta}
=(\sigma^m)_{\alpha\dot\beta}\partial_m\,.\lb{A1}
\ee
The same correspondence for the Maxwell field strengths is presented by the
relations
\bea
F_\alpha^\beta(A)&\equiv&
\frac14\partial_{\alpha\dot\beta}A^{\dot\beta\beta}
-\frac14\partial^{\dot\beta\beta}A_{\alpha\dot\beta}=
\frac18(\sigma^m\bar\sigma^n- \sigma^n\bar\sigma^m)_\alpha^\beta
F_{mn}\nn
&&=\,\frac18(\sigma^m\bar\sigma^n-
\sigma^n\bar\sigma^m)_\alpha^\beta F^+_{mn}\,,\lb{A2}\\
\bar{F}_{\dot\alpha}^{\dot\beta} &=&
\frac14(\bar\sigma^n)^{\dot\beta\beta}(\sigma^m)_{\beta\dot\alpha}F_{mn}
=-\frac18(\bar\sigma^m\sigma^n
-\bar\sigma^n\sigma^m)^{\dot\beta}_{\dot\alpha}F_{mn}\nn
&&=\,-\frac18(\bar\sigma^m\sigma^n
-\bar\sigma^n\sigma^m)^{\dot\beta}_{\dot\alpha}F^-_{mn}\,,\lb{A3}
\eea
where
\bea
&&F_{mn}=\partial_mA_n-\partial_nA_m,\quad
\tilde{F}_{mn}=\frac12\varepsilon_{mnrs}
F^{rs},\quad F^+_{mn}=\frac12F_{mn}+\frac{i}2\tilde{F}_{mn}\,, \\
&&\widetilde{F^+}_{mn}=-iF^+_{mn},\quad
F^-_{mn}=\frac12F_{mn}-\frac{i}2\tilde{F}_{mn}\,,\quad
\widetilde{F^-}_{mn}=iF^-_{mn}\,.\lb{A4}
\eea
Thus, $F_{\alpha\beta}$ is the equivalent bispinor notation for the self-dual tensor field
$F^+_{mn}\,$, and $\bar{F}_{\dot\alpha\dot\beta}$ amounts to the anti-self-dual tensor $F^-_{mn}\,$.

The scalar variables in the spinor formalism are related to the
analogous variables in the tensor formalism as
\bea
&&\varphi=
F^{\alpha\beta} F_{\alpha\beta}=t+iz=\frac12F^{+mn}F^+_{mn}\,,\nn
&&\bar\varphi=\bar{F}^{\dot\alpha\dot\beta}\bar{F}_{\dot\alpha\dot\beta}
=t-iz=\frac12F^{-mn}F^-_{mn}\,,\lb{A5}\\
&&
t=\frac14F^{mn}F_{mn} = \frac12(\varphi + \bar\varphi)\,,\quad z=\frac14F^{mn}\tilde{F}_{mn}
= \frac{1}{2i}(\varphi - \bar\varphi)\,,\lb{A5b}\\
&&\frac{\partial}{\partial\varphi}=\frac12\partial_t-\frac{i}2\partial_z\,,\quad
\frac{\partial}{\partial\bar\varphi}=\frac12\partial_t+\frac{i}2\partial_z\,.
\lb{A6}
\eea

The bispinor and tensor representations of the dual field strengths
appearing in the nonlinear equations of motion are related as
\bea
&&P_\alpha^\beta(F)=\frac18(\sigma^m\bar\sigma^n-
\sigma^n\bar\sigma^m)_\alpha^\beta G^+_{mn}\,,\quad
\bar{P}_{\dot\alpha}^{\dot\beta}=-\frac18(\bar\sigma^m\sigma^n-
\bar\sigma^n\sigma^m)_{\dot\alpha}^{\dot\beta} G^-_{mn}\,,\lb{A7}\\
&&G^\pm_{mn}=\frac12G_{mn}\pm \frac{i}2\tilde{G}_{mn}\,,\quad
\tilde{G}_{mn}= 2\,\frac{\Delta L} {\Delta F^{mn}}\,,\quad
G_{mn}=-\varepsilon_{mnrs}\frac{\Delta L} {\Delta F_{rs}}\,,
\eea
where we employed the Lagrange derivatives.

The similar relations are valid for the auxiliary fields
\bea
V_\alpha^\beta=\frac18(\sigma^m\bar\sigma^n-
\sigma^n\bar\sigma^m)_\alpha^\beta V^+_{mn}\,,\qquad
\bar{V}_{\dot\alpha}^{\dot\beta}=-\frac18(\bar\sigma^m\sigma^n-
\bar\sigma^n\sigma^m)_{\dot\alpha}^{\dot\beta} V^-_{mn}\,.\lb{A9}
\eea
The complex auxiliary tensorial fields $V^{\pm}_{mn}$ are expressed
through the real fields as
\bea
&&V^+_{mn}=\frac12V_{mn}+\frac{i}2\tilde{V}_{mn}\,,\quad V^-_{mn}=\frac12V_{mn}-\frac{i}2\tilde{V}_{mn}\,,\\
&&\widetilde{\tilde{V}}_{mn}=-V_{mn}\,,\quad
\tilde{V}^{mn}\tilde{V}_{mn} =-V^{mn}V_{mn}\,.\lb{A10}
\eea
The scalar
variable $\nu$ can be represented as
\bea
&&\nu=V^{\alpha\beta}V_{\alpha\beta}=v+iw\,,\qquad v=\frac14V^{mn}V_{mn}\,,\quad w=\frac14\tilde{V}^{mn}V_{mn}\,,\lb{A12} \\
&&\partial_\nu=\frac12\partial_v-\frac{i}2\partial_w\,.\lb{A11}
\eea

The real tensor and scalar fields just introduced have the following
$O(2)$ transformation laws
\bea
&&\delta_\omega V_{mn}=\omega\tilde{V}_{mn}\,,\quad \delta_\omega\tilde{V}_{mn}=-\omega V_{mn}\,,\\
&&\delta_\omega v=2\omega w\,,\quad \delta_\omega w=-2\omega v\,.\lb{A13}
\eea
In the tensorial notations our basic $O(2)$ invariant variable $a$ is
reduced to
\bea a=\nu\bar\nu=v^2+w^2\,.\lb{A14}
\eea

The one-to-one correspondence between the specific variables used in
\cite{BN,CKR,CKO} and our variables in the tensor notation is
as follows \footnote{Sometimes, for brevity, we omit the
antisymmetric tensor indices. }:
\bea
&&T=F-iG=F-2i\tilde{V}+i\tilde{F}\,,\quad
\tilde{T}=\tilde{F}-i\tilde{G} =\tilde{F}+2iV-iF\,, \nn
&&
T^*=F+iG=F+2i\tilde{V}-i\tilde{F}\,,\quad\widetilde{T^*}=
\tilde{F}-2iV+iF\,,\\
&&\delta_\omega T=i\omega T\,,\quad\delta_\omega \tilde{T}=i\omega
\tilde{T}\,,\quad\delta_\omega T^*=-i\omega T^*\,,\quad
\delta_\omega \widetilde{T^*}=-i\omega \widetilde{T^*}\,.
\eea
Here Eqs.\p{GFV} and \p{GFV2} were used.
The relations between self-dual (and anti-self-dual) parts of these
two sets of complex variables can be collected as
\bea
&&T^+=\frac12(T+i\tilde{T})=(F-V)+i(\tilde{F}-\tilde{V})
=2F^+-2V^+\,,\lb{A15}\\
&&
T^-=\frac12(T-i\tilde{T})=V-i\tilde{V}=2V^-\,,\lb{A16}\\
&&T^{*+}\equiv\bar{T}^+=(T^-)^* = \frac12(T^*+i\widetilde{T^*})
=V+i\tilde{V}=2V^+\,,\lb{A17}\\
&&T^{*-}\equiv\bar{T}^-=(T^+)^*=\frac12(T^*-i\widetilde{T^*})
=F-i\tilde{F}-V+i\tilde{V}=2F^--2V^-\,.\lb{A18}
\eea

Being cast in the tensor notations, our Lagrangian \p{LVF} becomes:
\bea
&&{\cal
L}(F,V)=-\frac14F^{mn}F_{mn}+\frac12(V^{mn}-F^{mn})(V_{mn}-F_{mn})+{\cal
E}(v^2+w^2)\nn
&&=-\frac14[(F^+)^2+(F^-)^2]+\frac12[(V^+-F^+)^2+(V^--F^-)^2] +{\cal
E}(v^2+w^2)\lb{TFV}
\eea
and $v^2+w^2 =\frac14(V^+)^2(V^-)^2\,$. The dual nonlinear field strength $\tilde{G}^{mn}$ is expressed through our
variables as
\bea
&&\tilde{G}^{mn}(F,V)=2\frac{\partial{\cal
L}(V,F)}{\partial F_{mn}}= (F -2V)^{mn}\,. \lb{GFV}
\eea
Applying the tilde operation to both sides of this equality yields
\be
 G^{mn}(F,V)=(2\tilde{V}-\tilde{F})^{mn}\,.\lb{GFV2}
\ee
In terms of self-dual components, the same relations read
\bea
&&G^+=\frac12(G+i\tilde{G})=iF^+-2iV^+\,,\qquad V^+=\frac12(F^++iG^+)\,,
\nn
&& G^-=\frac12(G-i\tilde{G}) =-iF^-+2iV^-\,,\qquad
V^-=\frac12(F^--iG^-)\,.
\eea

The auxiliary equations of our formalism \p{FVequ} are completely equivalent to the ``twisted
nonlinear self-duality condition'' of Refs. \cite{BN,CKR}.

The tensor version of our algebraic equation reads
\bea
(F^+-V^+)_{mn}=\frac{\partial{\cal E}(a)}{\partial
V^{+mn}}=\frac12V^+_{mn}(V^-)^2{\cal E}_a\,.\lb{ours}
\eea
After passing to the $T$-tensor notation by Eqs. \p{A16}, \p{A17}, the
same equation is rewritten as
\bea
T^+_{mn}=4\frac{\partial{\cal
E}(a)}{\partial \bar{T}^{+mn}}= \frac18(\bar{T}^+)_{mn}(T^-)^2{\cal
E}_a\,, \qquad
a=\frac14(V^+)^2(V^-)^2=\frac1{64}(\bar{T}^+)^2(T^-)^2\,,\lb{tensalg}
\eea
that precisely coincides with the general twisted self-duality
condition. Our interaction function ${\cal E}$ proves to be
identical to the  ``deformation function'' ${\cal I}^{(1)}$ used in
\cite{CKR,CKO}. The general case with derivatives on the Maxwell field
strengths corresponds to passing to the Lagrange derivatives in
\p{ours} and \p{tensalg}.

\setcounter{equation}0

\renewcommand{\theequation}{B.\arabic{equation}}
\section*{B. Schr\"odinger constraint and BI theory}
Similarly to \cite{GZ2,CKR}, we consider here the Schr\"odinger type \cite{Schr} nonlinear constraints
\bea
&&T_{mn}T^{rs}\tilde{T}_{rs}-\tilde{T}_{mn}T^{rs}T_{rs}=\pm\frac18
\tilde{T}^*_{mn}(T^{rs}\tilde{T}_{rs})^2.\lb{Schr}
\eea
The contraction of this condition with $T^{mn}$ yields the NGZ self-duality constraint
\bea
T^{mn}\tilde{T}^*_{mn}=F^{mn}\tilde{F}_{mn}+G^{mn}\tilde{G}_{mn}=0
\eea
for both signs in the right-hand side of \p{Schr}.

For definiteness we choose the sign minus and, using the relations from the Appendix A,
equivalently rewrite \p{Schr} in the bispinor notations as the two conditions
\bea
&&2(F-V)_{\alpha\beta}\bar{V}^2=V_{\alpha\beta}
[\bar{V}^2-(F-V)^2]^2\,, \nn
&& 2\bar{V}_{\dot\alpha\dot\beta}(F-V)^2
=(\bar{F}-\bar{V})_{\dot\alpha\dot\beta}[\bar{V}^2-(F-V)^2]^2\,.\lb{Schrsd}
\eea
Now we substitute the twisted self-duality equation \p{ours} in its spinorial form \p{FVequDu}  into
these constraints and find that the latter imply the following equation for the auxiliary interaction
\bea
&&2{\cal E}_a=[1-a{\cal E}_a^2]^2\,.\lb{eqEa}
\eea
This equation is greatly simplified in terms of the variable $b$ and the interaction $I(b)$
defined in \p{relab}:
\be
I_b = -\frac{2}{(b -1)^2} \; \rightarrow \; I(b) = \frac{2b}{b -1}\,.
\ee
It is recognized as the interaction defining the BI theory.

The sign plus in the constraint \p{Schr} corresponds to
the ``twisted'' solution $E(a)\rightarrow -E(a)$ which amounts to the choice of the opposite sign of
the BI coupling constant,  $f^2\rightarrow -f^2\,$. These two options exhaust all
possible solutions of the constraints \p{Schr}.

\setcounter{equation}0

\renewcommand{\theequation}{C.\arabic{equation}}
\section*{C. Manifestly duality-invariant action
with auxiliary ${\;\;\;\;\,}$ fields}
It is known that the duality-invariant
systems admit an alternative description in which the $O(2)$ duality
symmetry becomes manifest at cost of loosing the manifest Lorentz
symmetry (see, e.g., \cite{RT,RT2})\footnote{Both symmetries can be made
explicit after introducing some additional gauge degrees of freedom
\cite{PST}; here we do not touch this type of theories.}. Here we
demonstrate that the formulation with the auxiliary tensorial fields
can also be rearranged in a similar way.

We start by introducing the $3D$ notations for the Maxwell field
strength $F_{mn}(A)=\partial_mA_n-\partial_nA_m\,$ and the auxiliary
fields $V_{mn}$:
\bea
&&E_k(A) := F_{0k}(A) =
\partial_0A_k-\partial_kA_0\,,\quad \varepsilon_{klj}B_j := F_{kl}(A)
= \partial_kA_l-\partial_lA_k\,,\nn
&&\tilde{E}_{k}=B_k, \quad \tilde{B}_{k} = -{E}_{k}\,, \quad
\tilde{F}_{kl}=-\varepsilon_{klj}E_j\,,\\
&&V_k := V_{0k}\,,\quad \varepsilon_{klj}U_j := V_{kl}\,,
\eea
where
$k,l = 1,2,3\,$. In the new setting, the scalar combinations of the
auxiliary fields  are rewritten as
\bea
&&\nu=v+iw=\frac12U_kU_k-\frac12V_kV_k-iV_kU_k\,,\\
&&a(V,U)=v^2+w^2=\frac14(U_iU_i)^2+\frac14(V_iV_i)^2-\frac12(U_iU_i)(V_kV_k)+
(V_iU_i)^2\,.
\eea
Next, we introduce one more gauge potential
$A^\prime_m\,$, as well as an extra independent auxiliary field $N_{mn} =
-N_{nm}$, and rewrite our Lagrangian \p{LVF} with $E_(\nu, \bar\nu) = {\cal E}(a)$ in the equivalent form:
\bea
{\cal L}(F,V,A,
B)=\frac14\hat{F}^{mn}\hat{F}_{mn}-\hat{F}^{mn}V_{mn}+\frac12V^{mn}V_{mn}
-\frac14\varepsilon^{mnrs}N_{mn}F^\prime_{rs} +{\cal E}(v^2+w^2)\,,\lb{FVAB}
\eea
where $\hat{F}_{mn}=N_{mn}+ F_{mn}(A)$ and $F_{rs}^\prime
=\partial_r A^\prime_s-\partial_s A^\prime_r$. The field $A^\prime_m$
serves as the Lagrange multiplier producing the Bianchi identity for $N_{mn}\,$,
which can be solved as $N_{mn} = \partial_m C_n-\partial_n
C_m$. Then $\hat{F}_{mn}=\partial_m(C_n+A_n)-\partial_n(C_m+A_m)$,
which takes us back to the original Lagrangian \p{LVF}.

On the other hand, following \cite{RT}, we can keep both gauge
fields in the Lagrangian \p{FVAB}, in which case it displays the new
vectorial gauge symmetry
\be
\delta N_{mn} = \partial_m\lambda_n -
\partial_n\lambda_m\,, \quad \delta A_m = - \lambda_m\,.
\ee
We can make use of this gauge freedom to fix the magnetic non-covariant gauge
\be
N_{0k}=0\,.\lb{noncovG}
\ee
It leaves us with $N_{kl} =\varepsilon_{klj}N_j$ and implies
$$
\frac12N^{mn}N_{mn}= N_kN_k\,,\;
N^{mn}V_{mn}= 2N_kU_k\,,\;
\frac12\tilde{N}^{mn}F^\prime_{mn}= -N_kE^\prime_k\,,\;
\frac12N^{mn}F_{mn}=N_kB_k\,.
$$

In the gauge \p{noncovG}, the bilinear part of our Lagrangian reads:
\bea
{\cal L}_2(N,V,U,A,A^\prime)&=& \frac12N_kN_k-N_k(2U_k-B_k-E^\prime_k)
-\frac12 E_kE_k
+\frac12 B_kB_k\nn
&&+\,2 E_kV_k-2 B_kU_k- V_kV_k+ U_kU_k\,.\lb{noncov1}
\eea
After eliminating the auxiliary field $N_k$ by the equation of
motion,
\bea N_k=2U_k-B_k-E^\prime_k\,,
\eea
the whole $O(2)$ invariant Lagrangian with two gauge potentials and the auxiliary fields $V_k,
U_k$ is finally written as
\bea
{\cal L}(A,A^\prime,V,U)&=&-\frac12 E_kE_k-\frac12 E^\prime_kE^\prime_k+
 B_kE^\prime_k+2 E_kV_k
+2 U_kE^\prime_k\nn
&& -\, V_kV_k- U_kU_k+{\cal E}[a(V,U)]\,.\lb{VUAB1}
\eea

The $O(2)$ duality transformations are realized on these fields as
\bea
&&
\delta_\omega V_k=\omega U_k\,, \quad \delta_\omega
U_k=-\omega V_k\,,\nn
&&\delta_\omega E_k= \omega E'_k\,,\;\; \delta_\omega E'_k=-
\omega E_k\,, \quad \delta_\omega B_k= \omega B'_k\,,\; \delta_\omega B'_k= -\omega B_k\,.
\eea
Then, using the relations
$$
E_kB_k = -\frac14 F^{mn}(A)\tilde{F}_{mn}(A) = \mbox{div}\,,\quad
E'_k B'_k  = -\frac14 F^{mn}(A^\prime)\tilde{F}_{mn}(A^\prime)= \mbox{div}\,,
$$
we prove the $O(2)$ invariance of the non-covariant  action
\bea
&&\delta S=\int d^4x\delta L(A,A^\prime,V,U)=\omega\int d^4x (
E^\prime_kB^\prime_k-E_kB_k)=0\,.
\eea

The twisted self-duality
equations have the following form in this formalism:
\bea
&&V_k=E_k+\frac12{\cal E}_a\{V_k[(V_lV_l)-(U_lU_l)]+
2U_k(V_lU_l)\}\,,
\\
&&U_k=E^\prime_k+\frac12{\cal E}_a\{U_k[(U_lU_l)-(V_lV_l)]+
2V_k(V_lU_l)\}\,,\\
&&{\cal E}_a=\frac12+e_2a +\ldots\, .\nonumber
\eea
These algebraic equations can be solved
perturbatively in terms of the fields  strengths $E_k$ and $E'_k$. To the lowest order:
\bea
&&V_k=E_k+\frac14\{E_k[(E_lE_l)-(E^\prime_lE^\prime_l)]+
2E^\prime_k(E_lE^\prime_l)\}+O(E^5)\,,
\\
&&U_k=E^\prime_k+\frac14\{E^\prime_k[(E^\prime_lE^\prime_l)-(E_lE_l)]+
2E_k(E_lE^\prime_l)\}+O(E^5)\,.
\eea
 After substituting
these expressions back into \p{VUAB1}, we obtain the manifestly
self-dual Lagrangian as a function of the $3D$ field strengths
\bea
L^{sd}(A,A^\prime)&=& \frac12 E_kE_k+\frac12 E^\prime_kE^\prime_k+ B_kE^\prime_k+
\frac18( E_kE_k)^2+\frac18( E^\prime_kE^\prime_k)^2\nn
&&-\,\frac18( E_kE_k)( E^\prime_lE^\prime_l)+ \frac12( E_kE^\prime_k)^2+\ldots\,.
\eea
The magnetic version of the manifestly $O(2)$ invariant action can be obtained through
the discrete duality transformation of both fields
$E\rightarrow B,\quad B\rightarrow -E$.

 

\begin{thebibliography}{99}


\bibitem{GZ}M.K. Gaillard and B. Zumino, {\it Duality rotations for interacting
fields}, Nucl. Phys. B {\bf 193} (1981) 221.
\bibitem{GZ2}M.K. Gaillard and B. Zumino, {\it Self-duality in nonlinear
electromagnetism}, In: Supersymmetry and quantum field theory, eds.
J. Wess and V.P. Akulov, p. 121-129,
Springer-Vellag, 1998, {\tt hep-th/9705226};\\
M.K. Gaillard and B. Zumino, {\it Nonlinear electromagnetic
self-duality and Legendre transform}, In: Duality and Supersymmetric
Theories, eds. D.I. Olive and P.C. West, p. 33, Cambridge University
Press, 1999, {\tt hep-th/9712103}.
\bibitem{GR} G.W. Gibbons and D.A. Rasheed, {\it Electric-magnetic duality
rotations in nonlinear electrodynamics}, Nucl. Phys. B {\bf 454}
(1995) 185, {\tt hep-th/9506035}.
\bibitem{KT} S.M. Kuzenko and S. Theisen, {\it Nonlinear self-duality and
supersymmetry}, Fortsch. Phys. {\bf 49} (2001) 273, {\tt
hep-th/0007231}.
\bibitem{DT}S. Deser, C. Teitelboim, {\it Duality transformations of abelian
and nonabelian gauge fields}, Phys. Rev. D {\bf 13} (1976) 1592.
\bibitem{HT} M. Henneaux, C. Teitelboim, {\it Dynamics of chiral (selfdual) p-forms},
Phys. Lett. B {\bf 206} (1988) 650.
\bibitem{SS}J.H. Schwarz, A. Sen, {\it Duality symmetric actions}, Nucl. Phys.
B {\bf 411} (1994) 35, {\tt hep-th/9304154}.
\bibitem{HKS}
M. Hatsuda, K. Kamimura and S. Sekia, {\it Electric-magnetic duality
invariant Lagrangians}, Nucl. Phys. B {\bf 561} (1999) 341, {\tt
hep-th/9906103}.
\bibitem{BC} X. Bekaert and S. Cucu, {\it Deformations of duality-symmetric
theories}, Nucl. Phys. B {\bf 610} (2001) 433, {\tt hep-th/0104048}.
\bibitem{PST} P. Pasti, D. Sorokin and M. Tonin, {\it Duality symmetric actions
with manifest space-time symmetries}, Phys. Rev. D {\bf 52} (1995)
R4277, {\tt hep-th/9506109};\\
P. Pasti, D. Sorokin and M. Tonin, {\it Covariant actions for models
with non-linear twisted self-duality},  Phys.Rev. D {\bf 86} (2012)
045013,
 {\tt arXiv:1205.4243 [hep-th]}.
\bibitem{RT} M. Ro\v{c}ek and A. Tseytlin, {\it Partial breaking of global
D=4 supersymmetry, constrained superfields, and 3-brane actions},
Phys. Rev. D {\bf 59} (1999) 106001, {\tt hep-th/9811232}.
\bibitem{Ts} A.A. Tseytlin, {\it Born-Infeld action, supersymmetry and string
theory}, In: The many faces of superworld, Yu. Golfand memorial
volume, ed. M.A. Shifman, p. 417, World Scientific, 2000, {\tt
hep-th/9908105}.
\bibitem{ABMZ}P. Aschieri, D. Brace, B. Morariu, B. Zumino,
{\it Proof of a symmetrized trace conjecture for the abelian
Born-Infeld Lagrangian}, Nucl. Phys. B {\bf 588} (2000) 521, {\tt
hep-th/0003228}.
\bibitem{AFZ}P. Aschieri, S. Ferrara and B. Zumino, {\it Duality
rotations in nonlinear electrodynamics and extended supergravity},
Riv. Nuovo Cim. {\bf 31} (2008) 625, {\tt arXiv:0807.4039 [hep-th]}.
\bibitem{BHN} G. Bossard, C. Hillmann, H. Nicolai, {\it $E_{7(7)}$ symmetry in perturbatively
quantised ${\cal N}=8$ supergravity}, JHEP {\bf 1012} (2010) 052,
{\tt arXiv:1007.5472 [hep-th]}.
\bibitem{RKa} R. Kallosh, {\it $E_{7(7)}$ symmetry and finiteness of ${\cal N}=8$ supergravity}, {\tt arXiv:1103.4115 [hep-th]};
R. Kallosh, {\it ${\cal N}=8$ counterterms and $E_{7(7)}$ current
conservation}, JHEP {\bf 1106} (2011) 073, {\tt arXiv:1104.5480
[hep-th]}.
\bibitem{BN} G. Bossard, H. Nicolai, {\it Counterterms vs. dualities},
JHEP {\bf 1108} (2011) 074, {\tt arXiv:1105.1273 [hep-th]}.
\bibitem{CKR}J.J.M. Carrasco, R. Kallosh, R. Roiban, {\it Covariant procedure for
perturbative nonlinear deformation of duality-invariant theories},
Phys. Rev. D {\bf 85} (2012) 025007, {\tt arXiv:1108.4390 [hep-th]}.
\bibitem{CKO}W. Chemissany, R. Kallosh, T. Ortin, {\it Born-Infeld with higher derivatives},
Phys. Rev. D {\bf 85} (2012) 046002, {\tt arXiv:1112.0332 [hep-th]}.
\bibitem{RT2}R. Roiban, A. Tseytlin, {\it On duality symmetry in perturbative
quantum theory}, JHEP {\bf 1210} (2012) 099, {\tt arXiv:1205.0176
[hep-th]}.
\bibitem{IZ} E.A. Ivanov, B.M. Zupnik, {\it ${\cal N}=3$ supersymmetric Born-Infeld
theory}, Nucl. Phys. B {\bf 618} (2001) 3, {\tt hep-th/0110074}.
\bibitem{IZ1} E.A. Ivanov, B.M. Zupnik, {\it New representation for Lagrangians
 of self-dual nonlinear electrodynamics}, In: Supersymmetries and
quantum symmetries, eds. E. Ivanov et al, p. 235, Dubna, 2002, {\tt
hep-th/0202203}.
\bibitem{IZ2}
 E.A. Ivanov, B.M. Zupnik, {\it New approach to nonlinear
electrodynamics: dualities as symmetries of interaction}, Yadern.
Fiz. {\bf 67} (2004)  2212; [Phys.  Atom. Nucl.  {\bf 67} (2004)
2188], {\tt hep-th/0303192}.
\bibitem{Schr}
E. Schr\"odinger, {\it Contributions to Born's new theory of the electromagnetic field},
Proc. Roy. Soc. (London) A {\bf 150} (1935) 465.
\bibitem{UN}
E.A. Ivanov, B.M. Zupnik, {\it Bispinor auxiliary fields in duality-invariant
electrodynamics revisited:
    the $U(N)$ case}, in preparation.
\end{thebibliography}
\end{document}